\documentclass{article}

\usepackage{jheppub}
\usepackage[utf8]{inputenc}
\usepackage{multirow}
\usepackage[toc,page]{appendix}

\usepackage{amsmath}
\usepackage{amsfonts}
\usepackage{amssymb}
\usepackage{graphicx}
\usepackage{ulem} 
\usepackage{slashed}
\usepackage{hyperref}
\usepackage{float}
\usepackage{csquotes}

\usepackage{graphicx}
\usepackage{caption}
\usepackage{subcaption}

\usepackage[ruled,vlined]{algorithm2e}
\usepackage[most]{tcolorbox}

\usepackage{placeins}

\makeatletter
\gdef\@fpheader{\vline}
\makeatother

\author[a,b]{Sascha Caron}
\emailAdd{scaron@nikhef.nl}
\author[c]{Jos\'e Enrique Garc\'ia Navarro}
\emailAdd{Jose.Enrique.Garcia@ific.uv.es}
\author[c]{Mar\'ia Moreno Ll\'acer}
\emailAdd{Maria.Moreno@ific.uv.es}
\author[a,b]{Polina Moskvitina}
\emailAdd{p.moskvitina@nikhef.nl}
\author[a]{Mats Rovers}
\emailAdd{mats.rovers@ru.nl}
\author[c]{Adri\'an Rubio J\'imenez}
\emailAdd{adrian.rubio.jimenez@cern.ch}
\author[c]{Roberto Ruiz de Austri}
\emailAdd{rruiz@ific.uv.es}
\author[d]{Zhongyi Zhang}
\emailAdd{zhongyi@th.physik.uni-bonn.de}

\affiliation[a]{High Energy Physics, Radboud University Nijmegen, Heyendaalseweg 135, 6525 AJ Nijmegen, the Netherlands}
\affiliation[b]{Nikhef, Science Park 105, 1098 XG Amsterdam, the Netherlands}
\affiliation[c]{Instituto de Física Corpuscular, IFIC-UV/CSIC, Catedrático José Beltrán 2, E-46980 Paterna, Spain}
\affiliation[d]{The Bethe Center for Theoretical Physics, Bonn University, 53115 Bonn, Germany}

\title{Universal Anomaly Detection at the LHC: Transforming Optimal Classifiers and the DDD Method}

\abstract{In this work, we present a novel approach to transform supervised classifiers into effective unsupervised anomaly detectors. The method we have developed, termed Discriminatory Detection of Distortions (DDD), enhances anomaly detection by training a discriminator model on both original and artificially modified datasets. We conducted a comprehensive evaluation of our models on the Dark Machines Anomaly Score Challenge channels and a search for 4-top quark events, demonstrating the effectiveness of our approach across various final states and beyond the Standard Model scenarios.

We compare the performance of the DDD method with the Deep Robust One-Class Classification method (DROCC), which incorporates signals in the training process, and the Deep Support Vector Data Description (DeepSVDD) method, a well-established and well-performing method for anomaly detection. Results show that the effectiveness of each model varies by signal and channel, with DDD proving to be a very effective anomaly detector. We recommend the combined use of DeepSVDD and DDD for purely unsupervised applications, with the addition of flow models for improved performance when resources allow. 

Findings suggest that network architectures that excel in supervised contexts, such as the particle transformer with standard model interactions, also perform well as unsupervised anomaly detectors. We also show that with these methods, it is likely possible to recognize 4-top quark production as an anomaly without prior knowledge of the process. We argue that the Large Hadron Collider community can transform supervised classifiers into anomaly detectors to uncover potential new physical phenomena in each search.
}

\begin{document}

\maketitle

\clearpage

\section{Introduction}
The Standard Model (SM) of particle physics cannot describe the complete world.
It is scientifically accepted that phenomena such as dark matter, baryon asymmetry, neutrino masses, and the description of gravity require extensions beyond the SM and/or General Relativity.
This extension may give rise to new particles and/or effects that are visible in the data from particle physics experiments.

\textbf{Assuming a signal and supervised event classifiers:} The search for these effects is like looking for a needle in a haystack. The production of new particles generally have a small cross-section and have to be filtered out of an enormous number of background events. Classification algorithms have been used for this purpose since the 1980s or 1990s, see e.g. \cite{Peterson:1993nk,Stimpfl-Abele:1993xgr,Ametller:1996ri,Denby:1987rk}
and since 2014 \cite{Baldi:2014kfa}, more and more deep learning algorithms have been used.
Classification algorithms such as neural networks learn with labelled data (i.e. supervised) to distinguish signals and background events given physical features (such as the 4-vectors and object types) based on training data by adjusting weights to minimize a so-called loss function. Typically, the loss function is a binary cross entropy, that measures the difference of the true class probability distribution to the predicted class probability distribution.  
Minimizing this penalty function then leads to the classifier learning the ratio of probability distributions in the feature space between signal and background.
In recent years, deep learning architectures have been shown to provide better classifiers than the boosted decision trees widely used in high-energy physics. More recently, fully connected deep learning architectures have been overtaken by graph networks~\cite{Shlomi_2021} and even attention-based transformer architectures~\cite{qu2024particle}, inspired by their success in language models~\cite{DBLP:journals/corr/VaswaniSPUJGKP17}.

\textbf{Not assuming a signal and unsupervised anomaly detection:} However, the question arises \enquote{which signal is considered?}. 
As mentioned above, the exact signal of physics beyond the SM (BSM) is not known, so a signal-by-signal search may not be efficient. One search method that has gained more and more attention in recent decades is the systematic search for anomalies, i.e., the model-independent search for events that are not SM-like.
One method is to test numerous signal regions using automated procedures~\cite{D0:2000vuh,Abbott:2000gx,Abbott:2001ke,Abazov:2011ma,H1:2004rlm,H1:2008aak,CDF:2007iou,CDF:2008voc, Aaboud:2018ufy,Sirunyan:2020jwk}.
Other approaches use unsupervised (or weakly supervised) classifiers. Machine Learning (ML) algorithms are trained without a signal model to estimate the distribution of SM events. Unsupervised refers to no signal data being used to train the ML model. Many approaches use AutoEncoders, mapping data features $x$ to a code $z$ and back to $x'$, assuming poor reconstruction of anomalies. In extreme unsupervised cases, only the expected density of the Standard Model is needed~\cite{Caron:2021wmq}, derived from simulated events, direct data (assuming a small signal), or auxiliary measurements.
Unsupervised Deep Learning techniques to search for new physics are discussed in~\cite{Wozniak:2020cry,Knapp:2020dde,Dery:2017fap,Cohen:2017exh,Metodiev:2017vrx,DAgnolo:2018cun,Heimel:2018mkt,Farina:2018fyg,Komiske:2018oaa,Hajer:2018kqm,Cerri:2018anq,Collins:2018epr,Otten:2019hhl,Blance:2019ibf,Collins:2019jip,Sirunyan:2019jud,vanBeekveld:2020txa,Aad:2020cws,Nachman:2020lpy,Kasieczka:2021xcg,Caron:2021wmq,Aad_2024,CMS-PAS-EXO-22-026,Harris:2881089,Kosters:2022amb}. Comparisons of unsupervised approaches are in the Dark Machines challenge~\cite{Aarrestad_2022} using datasets to compare supervised strategies with unsupervised approaches. Another comparison of model-independent approaches using different black boxes based on data and expectation density comparisons is the LHC Olympics~\cite{Kasieczka:2021xcg}.

\text{\textbf{Contribution: Proposal to turn a Classifier into an Anomaly Detector:}} \\
In this study, a novel approach is introduced: transforming state-of-the-art supervised classifiers into anomaly detectors with minimal changes. As supervised classifiers, such as those based on graph networks and transformers with physical information, continue to improve, we aim to exploit these advances for anomaly detection. The objective is to find a simple method to utilize these classifiers, widely used at the Large Hadron Collider (LHC), for anomaly detection. This adaptation could potentially extend the applicability of anomaly detection across all analyses performed at the LHC, with minimal additional cost. Given that anomaly detectors are currently rare, this approach has the potential to significantly improve the ability to detect unexpected phenomena in LHC data. This article introduces a simple but effective strategy called Discriminatory Detection of Distortions (DDD) for detecting anomalies. With this approach, the structure of existing algorithms does not need to be changed. Instead, the current model is retrained with a distorted data set as a signal.

The effectiveness of the DDD strategy is compared against leading models from the Dark Machines competition, in particular the DeepSVDD model~\cite{deepSVDD:pmlr-v80-ruff18a} and a novel model known as Deep Robust One-Class Classification (DROCC), as outlined in Ref~\cite{goyal2020drocc}.
The models are evaluated on five different datasets: one is from the search for 4-top production at the LHC~\cite{ATLAS:2023ajo, CMS:2023ftu}, the other four from the Dark Machines Anomaly Score (DarkMachines) challenge.

This analysis assesses the performance of two different architectures when coupled with these methods. One is a simple multi-layer perceptron (MLP) and the other is the particle transformer model (ParT) \cite{qu2024particle}. 
Furthermore, an approach incorporating pairwise features and physical interactions in the attention matrix (ParT + SM interactions), identified as the best event classifier in Ref.~\cite{Builtjes:2022usj}, has been integrated into evaluation. This comprehensive evaluation provides insight into the effectiveness of the different architectures in conjunction with these anomaly detection techniques. 

Finally, it is important to emphasize that the proposed approaches, when implemented in experiments such as those at the LHC, do not require major changes to the systematic uncertainties or other aspects of the analysis chain. The statistical test, which checks whether the data are consistent with expectations for high output scores, would just be conducted twice. First, it would be applied to the results of the anomaly score (e.g., from methods like DDD or DeepSVDD to search for anomalies), and second, to the results of the classifier (to search for known signals).
In addition, the construction of the algorithm for recognising anomalies from the classifier only requires re-training (as with DDD).

The work is organized as follows: Section~\ref{sec: generation} describes the datasets, Section~\ref{sec: Architectures} outlines the machine learning architectures, Section~\ref{sec: Techniques} presents various techniques for transforming a supervised classifier into an anomaly detector, and Section~\ref{sec: Results} discusses the results.

\section{Datasets} \label{sec: generation}

Proton-proton collisions were simulated at a center-of-mass energy of 13 TeV. The hard scattering events were generated using \texttt{MG5\_aMC@NLO} version 2.7~\cite{Alwall_2014} with the \texttt{NNPDF31\_lo} parton distribution function set~\cite{NNPDF:2017mvq}, using the 5 flavor scheme. Parton showering was performed with \texttt{Pythia} version 8.239~\cite{Sj_strand_2015}, and the MLM merging scheme~\cite{Mangano_2003} and a merging scale of $Q_0$ = 30 GeV was employed to merge high-multiplicity hard scattering events with the parton shower. The fast detector simulation was performed using \texttt{Delphes} version 3.4.2~\cite{de_Favereau_2014} with a modified ATLAS detector map, and jet clustering is performed with FastJet~\cite{Cacciari_2012} using the anti-kt algorithm with a jet radius of 0.4. 


\subsection{The 4-top dataset}

The background processes involved are $t\bar{t} + X$, where $X = Z, W^+, W^+W^-, H$, and the signal processes involved the production of 4-top quarks. For details regarding the object preselection adopted, we refer to Ref.~\cite{Builtjes:2022usj}. In terms of object and event selection, we defined a signal region~\cite{ATLAS:2020hpj} that requires at least six jets, of which at least two are b-tagged, $H_T > 500$ GeV, and two leptons of the same sign or at least three leptons for each event. More details can be found in Ref.~\cite{Builtjes:2022usj}.

\subsection{The Dark Machines dataset}

The background processes included are listed in Table~2 of Ref. \cite{Aarrestad_2022} as well as the signals consisting of processes of new physics, such as supersymmetric processes with and without R-parity conservation as well as Z'-resonances (see Table~3 of Ref.~\cite{Aarrestad_2022} for more details). For details about the object preselection adopted, we refer to Ref.~\cite{Aarrestad_2022}. The event election criteria for the four channels of the DarkMachines dataset are detailed in Table~\ref{tab:channels}.

\begin{table}[h]
\centering
\begin{tabular}{l|c|c|c|c}
\textbf{Selection Criteria} & \textbf{Channel 1} & \textbf{Channel 2a} & \textbf{Channel 2b} & \textbf{Channel 3} \\
\hline
\hline
$H_T$ [GeV] & $\geq 600$ & - & $\geq 50$ & $\geq 600$ \\
\hline
$E_{T}^{miss}$ [GeV] & $\geq 200$ & $\geq 50$ & $\geq 50$ & $\geq 100$ \\
\hline
$E_{T}^{miss}/H_T$ & $\geq 0.2$ & - & - & - \\
\hline
$b$-jets with $p_T > 50$ GeV & $\geq 4$ & - & - & - \\
\hline
$b$-jet with $p_T > 200$ GeV & 1 & - & - & - \\
\hline
$N^{lep}$ (with $p_T^{lep} > 15$ GeV) & - & $\geq 3$ & $\geq 2$ & - \\
\end{tabular}
\caption{Selection criteria for the four DarkMachines channels.}
\label{tab:channels}
\end{table}

\subsection{Data Format} \label{sec: data}

The generated Monte Carlo data were saved as ROOT files and processed into CSV files using the described event selection. Each line in the CSV files contains low-level kinematical information like the four-momentum of each object as well as a label denoting the object type: jet (j), b-tagged jet (b), electron (e-) and positron (e+), muon (mu-) and anti-muon (mu+), and photon (g). In addition, the missing transverse energy and its azimuthal value are provided per event. 

The structure of the input dataset is chosen to be different for the MLP and the ParT models when exploring the DeepSVDD, DROCC, and DDD techniques. For the MLP case, the whole information is provided as an input layer with a fixed size. A cut in the maximum number of objects is applied in order to avoid a large contribution of zeros coming from the padding (up to 5 jets, 5 b-jets, 1 electron, 1 positron, 1 muon, and 1 antimuon). This limitation in the number of objects is not applied for the ParT architecture, since its power in establishing correlations through the attention mechanism should avoid any bias coming from such padding. The DeepSVDD and DROCC models can utilize both MLP and ParT networks for anomaly detection, adapting their input format accordingly. However, the DDD model exclusively employs transformers, allowing the ParT input format to be used without imposing limitations on the number of objects. Since we were adding and removing objects, it was not practical to adapt the MLP to the DDD approach.

\section{Machine Learning Architectures} \label{sec: Architectures}

\subsection{Multi-Layer Perceptron}

In order to interpret the success of a complex architecture performing an anomaly detection task, it is crucial to compare with simpler architectures playing the role of benchmarks \cite{goodfellow2016deep}. With this goal, a deep MLP implementation has been developed, as illustrated in Table~\ref{table:MLPs}.

\begin{table}[ht]
\centering
\begin{tabular}{c}
\textbf{MLP}\\
\hline \hline
Dense (size = 16, activation = relu) \\
BatchNormalization  \\ 
\hline
Dense (size = 8, activation = relu) \\
BatchNormalization  \\
\hline
\end{tabular}
\caption{MLP architecture with two hidden layers.}
\label{table:MLPs}
\end{table}

The MLP consists of two hidden layers, each followed by a batch normalization layer. The first hidden layer has 16 neurons with ReLU activation, and the second hidden layer has 8 neurons with ReLU activation. This simple architecture is intended to provide a baseline for performance comparison.

\subsection{Particle Transformer} 

The Particle Transformer (ParT) \cite{qu2024particle} is a transformer-based architecture designed for jet tagging, inspired by techniques from natural language processing~\cite{DBLP:journals/corr/VaswaniSPUJGKP17}. The ParT processes four-momentum vectors of the different particles in the event and auxiliary variables. 

The architecture includes multi-head attention mechanisms with eight Particle Attention Blocks and two Class Attention Blocks. In Particle Attention Blocks, the multi-head attention mechanism incorporates an interaction matrix $U$ to emphasize particle interactions. This matrix $U$ is designed to include both pairwise features and the SM interaction matrix with running coupling constants~\cite{Builtjes:2022usj}, which depend on the energy scale of the interaction, thereby capturing the dynamics of particle interactions more accurately. This variation of the self-attention mechanism is defined as:

\begin{equation} \label{eq:part-attention}
    \text{Attention}(Q,K,V) = \text{SoftMax}\left(Q K^T / \sqrt{d} + U\right) V
\end{equation}
where $Q$, $K$, and $V$ are $d$-dimensional linear projections of the particle embedding based on the input embedding or the output of the previous layer.

The application of the attention mechanism correlates every particle with all others, allowing the model to handle varying numbers of particles and maintain permutation equivariance. Classification is achieved by appending a classification token to the particle embeddings before the final layers of the transformer. This token interacts with the embeddings and is processed by a fully connected network (FCN), which also receives $E_T^{\text{miss}}$ and $E_{T}^{\mathrm{miss}} \phi$ to produce the classification label \cite{jet-universe}.

Pairwise features, specifically the logarithms of $\Delta R_{ij}$ and $m_{ij}^2$, were selected based on previous work \cite{Builtjes:2022usj}. Here, $\Delta R_{ij}$ measures the angular distance between two particles in the detector's pseudo-rapidity ($\eta$) and azimuthal angle ($\phi$) space, while $m_{ij}^2$ is the invariant mass squared of two particles. The logarithm of these features is taken for better numerical handling and improved learning efficiency.

Additionally, the implementation of this work uses the third interaction matrix, referred to as the SM interaction matrix (SM matrix[3]), as described in the previous studies \cite{Builtjes:2022usj}. This matrix incorporates energy-dependent coupling constants based on the Standard Model of particle physics. The running of the fine-structure constant $\alpha$ and the strong coupling constant $\alpha_s$ are calculated using their respective Renormalization Group Equations (RGE). By including these running coupling constants, the SM interaction matrix provides a structured approach to encoding the strengths of physical interactions in the ParT, enhancing its ability to learn and generalize from the data.

Two versions of the ParT model are explored: one that incorporates both pairwise features and the SM interaction matrix (ParT+SM), and another that does not include these features (ParT). Despite these differences, the main characteristics of the architecture are consistent across both versions. This is done to ensure that performance differences arise exclusively from the training techniques rather than a particular set of hyperparameters. 

Hyperparameters such as learning rates and batch sizes are adapted for each training technique to optimize performance. Table \ref{tab: ParT hyperparameters} summarizes the Particle Transformer hyperparameters common for both versions considered in this study. 

\begin{table}[ht]
\centering
\begin{tabular}{lc}
\textbf{Hyperparameter} & \textbf{Value} \\
\hline \hline
Embed MLP dimensions & $[128, 512, 128]$ \\
Pair embed MLP dimensions & $[64, 64, 64]$ \\
Number of attention heads & 8 \\
Number of layers & 8 \\
Number of class attention blocks & 2 \\
FC dimensions &  [64, 256, 64] \\
Auxiliary FC dimensions & [32, 32, 128] \\
\hline
\end{tabular}
\caption{Hyperparameters of the Particle Transformer.}
\label{tab: ParT hyperparameters}
\end{table}

\section{Techniques for converting a supervised classifier into an anomaly detector}
\label{sec: Techniques}

The previously described architectures were originally developed for supervised classification tasks. This section presents three methods for adapting these architectures for anomaly detection.

\subsection{DeepSVDD: Deep Support Vector Data Description  }
\label{sec:svdd}

DeepSVDD is recognized as a well-established method for anomaly detection \cite{deepSVDD:pmlr-v80-ruff18a,Caron:2021wmq,Kosters:2022amb,CrispimRomao:2020ucc}. The DeepSVDD method was identified as one of the best performing methods in the DarkMachines challenge, outperforming autoencoder approaches in particular
\cite{Aarrestad_2022}.
However, the use of complex architectures in combination with this technique remains unexplored. 

DeepSVDD is a one-class classifier that predicts whether an input sample belongs to the normal class by mapping it to a hypersphere in the feature space and measuring its distance from the center of this hypersphere.
The adaptation of such architectures involves the integration of  an additional output layer, the dimension of which is a new hyperparameter of the technique. Training focuses on accurately predicting a particular point of the output space, which itself is another hyperparameter. The effectiveness of the training procedure is measured by the loss function, which quantifies the distance of events from this point. Geometrically, the training performs a mapping of the background events around this central point, with events that deviate significantly from this center being classified as anomalies.

There are infinite possibilities for the choice of a centre in the output space, but the option taken in this study is proposed in Ref.~\cite{deepSVDD:pmlr-v80-ruff18a}. The center is computed as the mean of the network representations that results from an initial forward pass on the training data. Removing all bias terms from the network is also needed to avoid trivial solutions. In addition to these considerations, not making use of dropout has been shown to be a convenient approach, improving significantly the stability of the training. 

DeepSVDD methods can typically address this problem with different sets of weights, representing different minima in the weight space. However, some of these minima may lead to `sphere collapse', where the model predicts the output independently of the input. To enhance the stability of this method, an ensemble approach, as proposed in Ref.~\cite{Caron:2021wmq}, is used. This involves computing the mean of the scores from multiple models.
Separate models are trained with 2, 4, 8, and 16 output dimensions, respectively, and these models are then combined to form the ensemble.

Any classifier can be transformed into an anomaly detector taking into account the considerations described above. The procedure is summarized in Algorithm~\ref{alg:deepsvdd}.


\begin{algorithm}[H]
\caption{Transforming a Classifier into an Anomaly Detector using the DeepSVDD technique}
\label{alg:deepsvdd}
\SetAlgoLined
\textbf{Step 1:} 
Remove all bias terms from the network\;
\textbf{Step 2:} Avoid the use of dropout layers\;
\textbf{Step 3:} Add an output layer with a certain number of nodes, which will define the dimension of the output hyperspace\; 
\textbf{Step 4:} Re-define the loss function as the distance to a target center in the hyperspace, which needs to be minimized during the training on the background events\; 
\textbf{Step 5:} Define the center of the hypersphere as the mean of the output from an initial forward pass on the training data\; 
\textbf{Step 6:} To improve stability, train separate models with different output dimensions (e.g., 2, 4, 8, and 16 nodes in this case). Take the mean scores for each event. 
\end{algorithm}


\subsection{DROCC: Deep Robust One-Class Classification}
\label{sec:drocc}

DROCC \cite{goyal2020drocc} \footnote{To the best of our knowledge, this is the first application of the DROCC method in the context of LHC physics.} is a one-class classification method that assumes data from the target class lies on a well-sampled, locally linear, low-dimensional manifold. This assumption enables DROCC to generate synthetic outliers during training to robustly learn a boundary around the target class data. The main idea is to train a classifier that not only accurately classifies the given `normal' data but also is robust enough to classify unseen and potentially very different `anomalous' data as being out-of-class.

This is done through the loss function, which includes a component that actively searches for adversarial examples near the `normal' data points. These adversarial examples are synthetic data points that are slightly perturbed versions of the normal data and are hypothesized to lie close to the decision boundary between normal and anomalous classes. The objective is to ensure that the model can correctly classify these challenging examples as anomalies, thereby enhancing its robustness to real anomalous data it may encounter. In this way, avoids the phenomenon of `sphere collapse'.

In the proposed approach, training is conducted entirely with background (target class) events to compute the cross-entropy loss and the adversarial loss. Signals are introduced only during the validation phase, where they are used to compute evaluation metrics, such as the Area Under the Curve (AUC). These signals, however, are not part of the training process and are distinct from those used in the final evaluation after training. The signals used for validation are taken from Ref. \cite{Caron_2023}, whereas those employed during the final test phase are from Re. \cite{Aarrestad_2022}. This weak supervision approach is motivated by the instability observed in the AUC during training. Monitoring the AUC after each epoch reveals significant fluctuations, and stopping training arbitrarily can result in poor performance at the final evaluation. By incorporating signals in the evaluation phase, we stabilize the process and guide hyperparameter optimization, ensuring that the model converges to a more robust solution. This allows the DROCC method to remain a one-class classification technique during training, while still benefiting from signals in the validation phase to improve AUC stability and overall model performance.

Algorithm~\ref{alg:drocc} provides a step-by-step outline of the DROCC method.



\begin{algorithm}[H]
\caption{DROCC for Anomaly Detection. For more details, see \cite{goyal2020drocc}.}
\label{alg:drocc}

\textbf{Training Phase:} \\
\For{$\text{epoch} = 1$ to $\text{Max\_epochs}$}{
    \For{$\text{batch}$ in $X_{\text{background}}$}{
        Generate adversarial examples $X_{\text{adv}}$ for $X_{\text{bkg}}$: \\
        \quad Apply small random perturbations to $X_{\text{bkg}}$\;
        \quad Project perturbed samples onto a $\text{Hypersphere}$ constraint\;
        Compute loss: \\
        \quad $L_{\text{original}} = \text{CrossEntropy}(f_{\text{model}}(X_{\text{bkg}}), \text{BackgroundLabel})$\;
        \quad $L_{\text{adv}} = \text{CrossEntropy}(f_{\text{model}}(X_{\text{adv}}), \text{AnomalyLabel})$\;
        \quad $\text{TotalLoss} = L_{\text{original}} + L_{\text{adv}}$\;
        Update model parameters $\theta$ using gradient descent to minimize the $\text{TotalLoss}$\;
    }
    \textbf{Validation Phase:} \\
    Evaluate $f_{\text{model}}$ on validation set $(X_{\text{val\_bkg}}, X_{\text{val\_sig}})$\;
    Compute the AUC to monitor overfitting and guide early stopping\;
}

\textbf{Testing Phase:} \\
Evaluate $f_{\text{model}}$ on test set $(X_{\text{test\_bkg}}, X_{\text{test\_sig}})$\;
Compute final metrics.

\end{algorithm}

\noindent

\subsection{DDD: Discriminatory detection of distortions} 
\label{sec:ddd}

This subsection introduces a novel approach called the Discriminatory Detection of Distortions (DDD) method, which aims to enhance anomaly detection by training a discriminator model on both original and artificially modified (background) datasets.
Most importantly, the DDD method requires no modification to the classifier. The shift from a classifier to an anomaly detector is achieved simply by changing the signal training data to a distorted version of the background data.

The DDD method advances traditional anomaly detection techniques by integrating a variety of data modification strategies:

\begin{enumerate}
    \item \textbf{Data Shifting}: Kinematic variables are adjusted using a normal distribution centered at 1 with a certain standard deviation ($std$). The shifting is applied to energy $E$, pseudorapidity $\eta$, and azimuthal angle $\phi$. The transverse momentum $p_T$ is recalculated based on the modified energy $E$ and pseudorapidity $\eta$ to preserve the mass of the particle. For $E_{T}^{\mathrm{miss}} \phi$ values are kept within $[-\pi, \pi]$, and for $E_T^{\text{miss}}$, values are kept positive.
    \item \textbf{Object Addition}:  Random addition of new objects from other background events, including object type and 4-vector with a certain probability $p$.
    \item \textbf{Object Removal}: Random removal of existing objects with the same specified probability $p$.
\end{enumerate}

This combination was determined to be the most effective in improving model performance compared to other combinations such as shifting + adding or shifting + removing objects alone.

The primary objective is to train the discriminator model with new training sets, making it capable of accurately distinguishing between normal and distorted background data. However, if the background data is distorted too much, distinguishing between the background and the distorted background becomes trivial. Conversely, if the data is distorted too little, the task becomes impossible. To find an optimal balance, the typical AUC achieved in supervised signal training is employed as a metric to determine when the background is optimally distorted. In this study, an AUC value of 0.85 is targeted, as this threshold was found based on previous supervised searches, as described in Ref.~\cite{Builtjes:2022usj}. 
This is an important difference to other methods that distort the data to find anomalies, e.g. ~\cite{Dillon:2023zac,Favaro:2023xdl}.

As with the deepSVDDs, an ensemble of the four best models that come closest to the target AUC value of 0.85 was used to stabilise the method. The final metrics are calculated based on the average predictions of these four models.

The inputs and outputs of the DDD method can be summarized as follows: 
\subsection*{Input}
\begin{itemize}
    \item Background dataset $D_{\text{background}}$: A dataset containing only normal, unmodified data.
    \item Hyperparameter $std$: Used for data modification via a normal distribution centered at 1 with standard deviations ranging from 0.00 to 0.10.
    \item Probability range $p$: The probability for randomly adding and removing objects within each event, set between 1\% to 10\%.
\end{itemize}

\subsection*{Output}
\begin{itemize}
    \item Trained Discriminator Model ($M_{\text{discriminator}}$).
    \item AUC score: A performance metric indicating how well the model differentiates between normal and modified (distorted) data.
\end{itemize}

The DDD method is summarized in Algorithm~\ref{alg:ddd}.

\begin{algorithm}[H]
\caption{Discriminatory Detection of Distortions with Dynamic Data Modification}
\label{alg:ddd}
\SetAlgoLined

\textbf{Step 1: Data Preparation} \\
Load the background dataset $D_{\text{background}}$\; 
Split $D_{\text{background}}$ into two sets: $D_{\text{original}}$ (e.g., even events) and $D_{\text{mod\_source}}$ (e.g., odd events)\; 

\textbf{Step 2: Data Modification} \\
\For{each event in $D_{\text{mod\_source}}$}{
    Apply the modification function $f_{\text{mod}}(\theta)$, where $\theta \sim N(1, \text{std})$, with std ranging from 0.00 to 0.10, to generate the modified dataset $D_{\text{modified}}$\; 
    \For{each object in the event}{
        With probability $p$, randomly decide to add a new object\; 
        With the same probability $p$, randomly decide to remove an existing object\; 
    }
    Adjust $\theta$ within the specified standard deviation range for subsequent modifications\; 
}

\textbf{Step 3: Discriminator Model Definition} \\
Define the discriminator model $M_{\text{discriminator}}$\; 

\textbf{Step 4: Training} \\
Combine $D_{\text{original}}$ and $D_{\text{modified}}$ into a combined dataset $D_{\text{combined}}$\; 
Label $D_{\text{original}}$ as `0' and $D_{\text{modified}}$ as `1'\; 
Train $M_{\text{discriminator}}$ on $D_{\text{combined}}$\; 

\textbf{Step 5: Evaluation} \\
Calculate the AUC score using predictions on $D_{\text{original}}$ and $D_{\text{modified}}$\; 

\textbf{Step 6: Hyperparameter Adjustment} \\ 
\If{AUC score deviates significantly from the typical signal value (0.85 as mentioned)}{
    Adjust $\theta$ accordingly\; 
    Repeat the process from Step 2 until the desired AUC score is achieved. 
}
\end{algorithm}

\begin{figure}[htb]
    \centering
    \includegraphics[width=1.\textwidth]{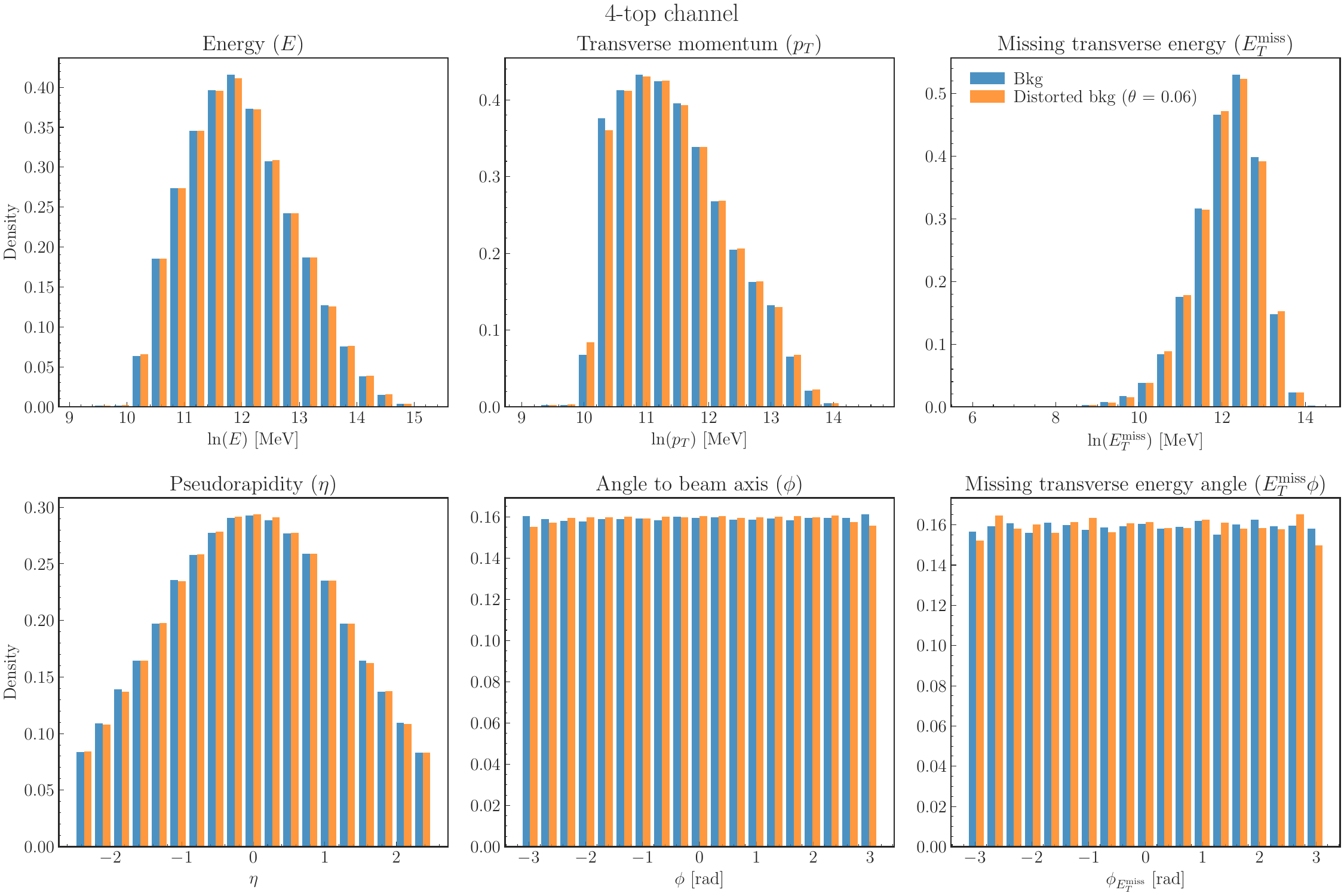}
    \caption{Distribution of various kinematic variables for the 4-top channel. The histograms compare the unmodified background (Bkg) and distorted background (Disctorted bkg) with $\theta$ = 0.06 and $p$ = 6\%. The variables Energy ($E$), Transverse momentum ($p_T$), and Missing transverse energy ($E_{T}^{\mathrm{miss}}$) are shown in MeV and in the log scale. Pseudorapidity ($\eta$), Angle to beam axis ($\phi$), and Missing transverse energy angle ($E_{T}^{\mathrm{miss}} \phi$) are also presented. Each histogram is normalized to have an area of 1, allowing shape comparisons across different datasets.}
    \label{fig: kinematic_distr}
\end{figure}

\begin{figure}[h!]
    \centering
    \begin{minipage}{0.65\textwidth}
        \centering
        \includegraphics[width=\textwidth]{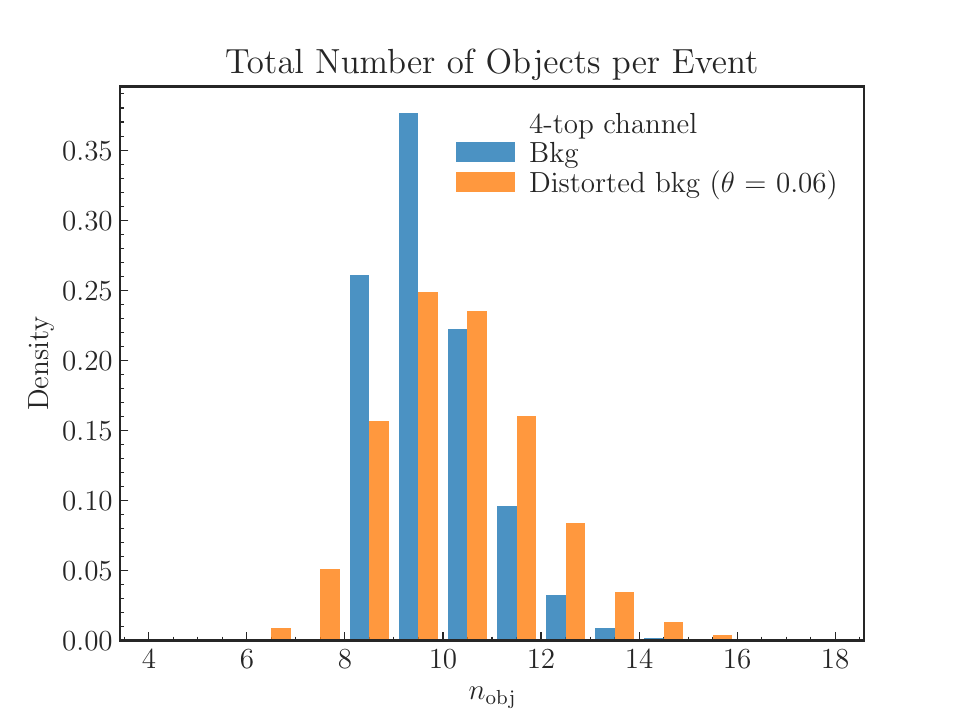}
    \end{minipage}
    \caption{Distribution of the total number of objects per event for the 4-top channel.}
    \label{fig: total_objects}
\end{figure}

The modification Function $(f_{\text{mod}})$ is designed to simulate potential anomalies by dynamically distorting (i.e. randomly adding and removing objects from the dataset) in addition to shifting data values.
Various hyperparameter $\theta$ and probability $p$ were tested. For each probability value and  $\theta$ the model is trained to find the best performing values until a good AUC is reached. Adjusting the hyperparameter $\theta$ is crucial for balancing the model's sensitivity to anomalies and ensuring robust performance across different conditions. Figure~\ref{fig: kinematic_distr} shows an example of the distorted and normal background distributions for a hyperparameter set used in the ensemble. Figure~\ref{fig: total_objects} demonstrates the distribution of the total number of objects per event. 

A strength of the DDD method lies in its general approach to data modification. By simultaneously incorporating data shifting, object addition and object removal, the method effectively improves the discriminator model's ability to recognise anomalies.
Furthermore, it is important to emphasize once again that the application of the method does not require a change in the network structure used for supervised signal-against-background training, which is commonly used in many HEP analyses. Various distortions of the background data can be analysed - on a case-by-case basis.  The method also uses ratios of probability densities and is invariant to variable transformations, which may be advantageous for some applications~\cite{Kasieczka:2022naq}.

\section{Results}
\label{sec: Results}

The performance of the MLP and ParT architectures, combined with the DeepSVDD, DROCC and DDD techniques, was evaluated on several data sets: the search for 4-top events (4-top channel) and the DarkMachines challenge channels~\cite{Aarrestad_2022}. The results are presented in detail below.

\FloatBarrier
\begin{figure}[h!]
    \centering
    \begin{subfigure}[b]{0.52\textwidth}
        \includegraphics[width=0.98\textwidth]{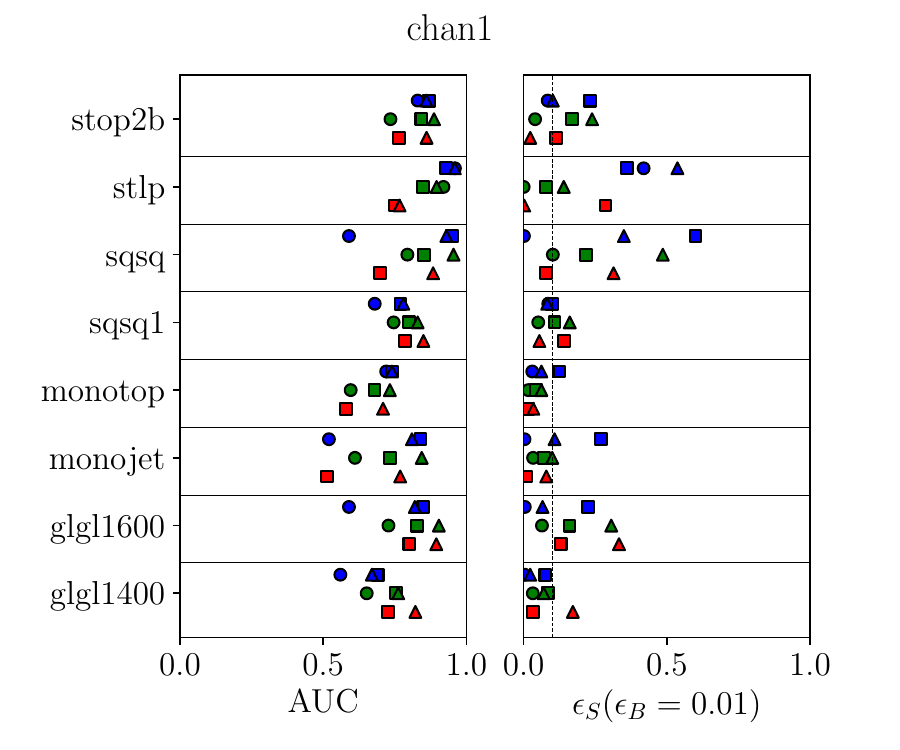}
    \end{subfigure}
    \hspace{-2.5em}
    \begin{subfigure}[b]{0.52\textwidth}
        \includegraphics[width=0.98\textwidth]{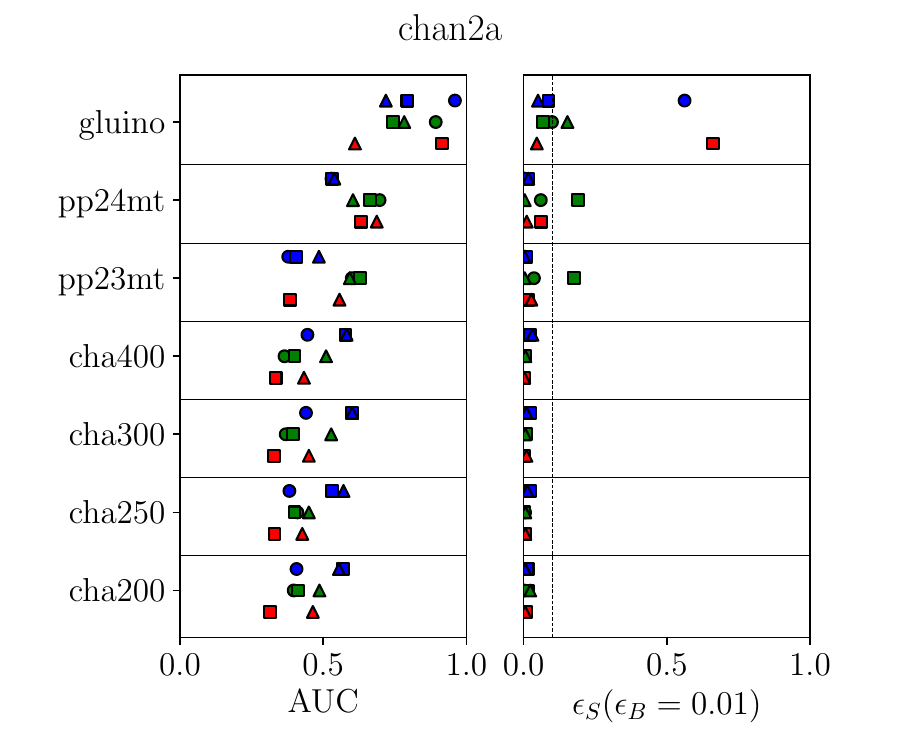}
    \end{subfigure}
    \vspace{-1em}
    \begin{subfigure}[b]{0.52\textwidth}
        \includegraphics[width=0.98\textwidth]{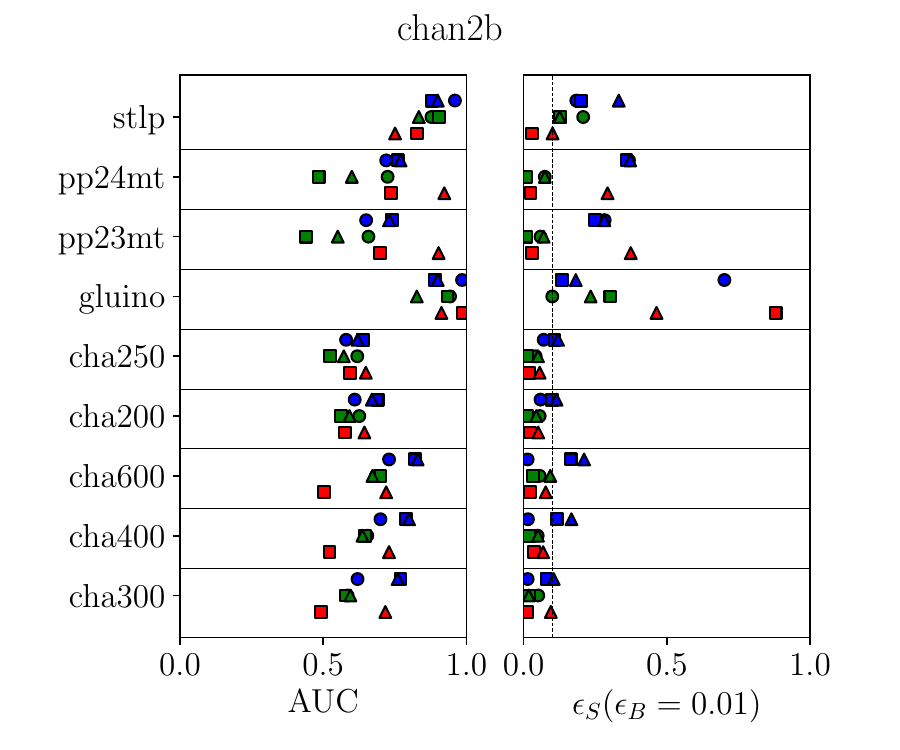}
    \end{subfigure}
    \hspace{-2.5em}
    \begin{subfigure}[b]{0.52\textwidth}
        \includegraphics[width=0.98\textwidth]{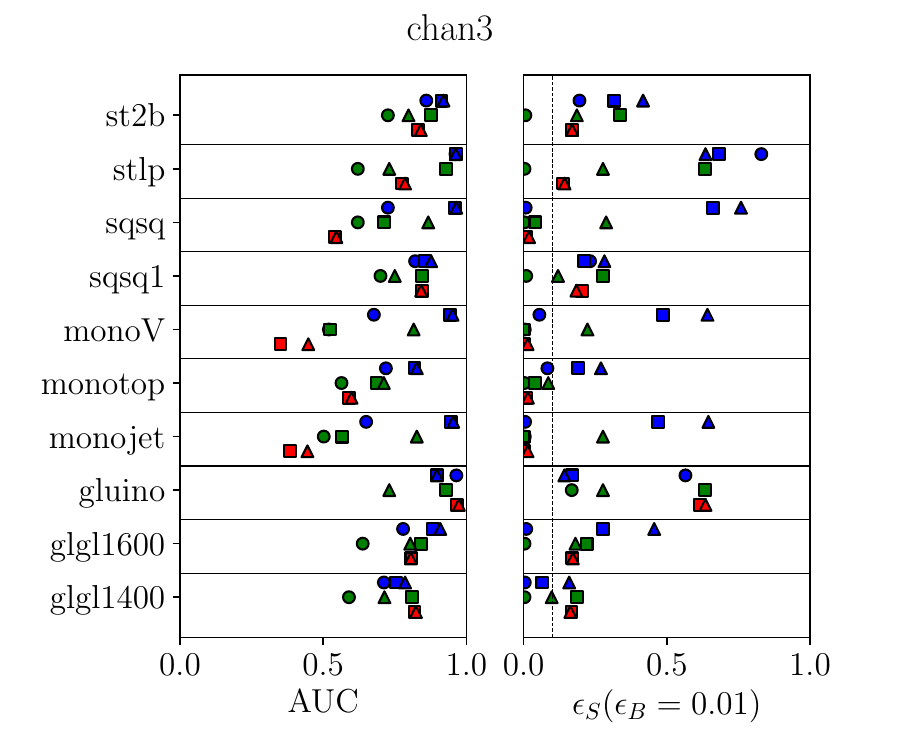}
    \end{subfigure}
    \\
    \vspace{2em}
    \begin{subfigure}[b]{0.75\textwidth}
        \centering
        \includegraphics[width=\textwidth]{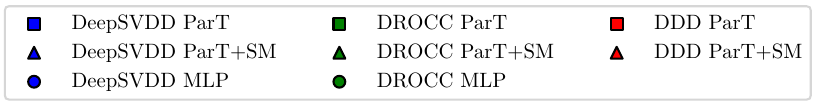}
    \end{subfigure}
    \caption{This figure shows the performance of the eight models on the four channels of the DarkMachines dataset: Channel 1 (first row, left), Channel 2a (first row, right), Channel 2b (second row, left) and Channel 3 (second row, right). For each channel, different signals are evaluated. The performance metrics shown are the AUC and the signal efficiency ($\epsilon_S$) assuming a background efficiency ($\epsilon_B$) of 1\%. A vertical line is added to indicate a significance improvement equal to 1. A color code is used to distinguish among the three training techniques: DeepSVDD (blue), DROCC (green) and DDD (red). Different shapes are used for each architecture: a circle for the MLP, a square for the ParT with no interactions, and a triangle for the ParT with SM couplings.} 
    \label{fig:comparisons_DarkMachines}
\end{figure}
\FloatBarrier

\subsection{Performance for the DarkMachines challenge channels}

In this subsection, the performance of the models trained for this study is evaluated for the DarkMachines challenge channels. This comprehensive evaluation helps to understand the effectiveness of these methods in a variety of final states and BSM scenarios.

Figure~\ref{fig:comparisons_DarkMachines} shows the results in such a way that a direct comparison with those from the DarkMachines publication is possible. The signal efficiency for a background efficiency setting of 1\% and the AUC are shown.
Different symbols represent different architectures: a circle for MLP, a square for ParT without interactions, and a triangle for ParT with SM couplings.

The observed trends are remarkably similar. The results for the individual channels are summarized as follows (numerical details of the measured variables used in this work can be found in Appendix~\ref{appendix:dmachinesTables}): 

\begin{itemize}

\item \textbf{Channel 1} exhibits high AUC values across most models, indicating strong anomaly detection capabilities. The ParT with pairwise interactions, including SM couplings, demonstrates superior performance in this channel. This highlights the ParT’s ability to handle complex correlations effectively. While the MLP shows competitive results, it generally lagged behind the more sophisticated ParT architectures, suggesting that simpler architectures might not capture the intricate patterns as well as the advanced ones.  Most of the signals from Channel 1 are characterized by large $E_T^{miss}$ and $p_T$ of the jets compared with the background. This suggests that the presence of these high-energy objects contributes to the strong performance observed across all models. 

\item \textbf{Channel 2a} is characterized by lower statistics, presents more challenges. Despite this, the ParT models continued to outperform the MLP. This indicates that while the ParT can still perform well with limited data, the inherent complexity of the dataset impacts the overall performance metrics. In this leptonic channel, $E_T^{miss}$ is less relevant than in Channel 1, with the $p_T$ of leptons playing an important role in discriminating signal from background.

\item \textbf{Channel 2b} shows similar trends to Channel 1 are observed, with robust performance metrics. The ParT models shows the highest sensitivity in this channel as well. This channel shares similarities with Channel 2a, with leptonic features being key discriminators.

\item \textbf{Channel 3} contains the largest dataset and provided the best overall performance metrics. This channel enables the models to fully utilize the available data for training and validation. The extensive dataset in this channel allows the models to generalize better and detect anomalies with higher precision, demonstrating the importance of a large amount of training data for improving model performance. Similar to Channel 1, $E_T^{miss}$ and the $p_T$ of jets are the most discriminative quantities, indicating a hadronic channel.
\end{itemize}

In summary, the DeepSVDD method shows consistent performance across all channels. It was particularly effective in Channel 3, where it achieves competitive AUC values. The MLP architecture with DeepSVDD shows solid performance in channels with less complex data, suggesting that it is suitable for simpler anomaly detection tasks where the data does not exhibit highly complex patterns.

The DROCC technique proves to be robust, especially with the ParT. It achieves high AUC values in all channels, especially in Channels 1 and 3. The ability of the DROCC method to generate synthetic outliers during training helped to create a robust decision boundary and improve the model's anomaly detection capabilities. The robustness of this method is evident in its consistent performance, making it a valuable technique for various anomaly detection scenarios.

The comprehensive approach of the DDD method, which includes data shifting, object addition and object removal, has proven to be effective in training discriminators to distinguish between normal and anomalous data. This strategy allows the DDD method to achieve competitive AUC and signal efficiency metrics, demonstrating its potential for advanced anomaly detection tasks.

Analysis of the DarkMachines challenge channels indicates that the ParT architecture consistently outperforms the MLP, particularly when pairwise interactions and SM couplings are included. Additionally, the choice of anomaly detection technique plays a crucial role in performance.


\subsection{Performance for the 4-top channel}
\label{subsec: 4top}

\begin{figure}[t!]
    \centering
    \begin{minipage}{0.65\textwidth} 
        \centering
        \includegraphics[width=\textwidth]{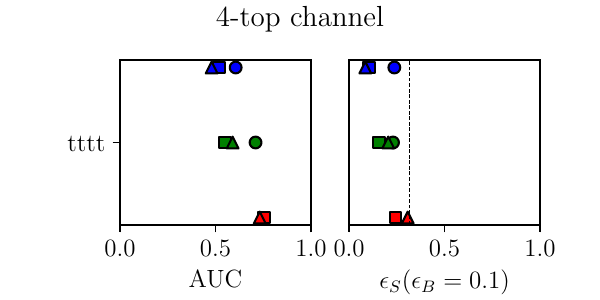}
    \end{minipage}
    \\
    \vspace{2em}
    \begin{minipage}{0.80\textwidth} 
        \centering
        \includegraphics[width=\textwidth]{Results/legend.pdf}
    \end{minipage}
    \caption{Performance of the eight models on the 4-top channel. The performance metrics shown are the AUC and the signal efficiency ($\epsilon_S$) assuming a background efficiency ($\epsilon_B$) of 10\%. A vertical line is added to indicate a significance improvement equal to 1. A color code is used to distinguish among the three training techniques: DeepSVDD (blue), DROCC (green), and DDD (red). Different shapes are used for each architecture: a circle for the MLP, a square for the ParT with no interactions, and a triangle for the ParT with SM couplings. }
    \label{fig:comparison_4tops}
\end{figure}

This subsection is dedicated to evaluate the performance of these techniques on the detection of events with 4-top quarks in the final state, which is a challenging signature predicted by the Standard Model with a very low cross-section that is possible to be produced at the LHC \cite{AguilarSaavedra:2011ug}. 
The idea behind this is: could 4-top production be detected as an anomaly without prior knowledge of its existence, i.e., would 4-top production be detected as an anomaly?

From Figure~\ref{fig:comparison_4tops} it is observed that the DDD method with ParT architecture (red triangle) achieves the highest AUC value (0.754) and a significant signal efficiency ($\epsilon_S$) of 0.307 at $\epsilon_B$ = 0.1. This highlights the strength of the ParT architecture in handling complex correlations through its attention mechanism, which effectively avoids bias from padding.

The DROCC technique also shows robust performance, particularly with the MLP architecture, achieving an AUC of 0.71 and $\epsilon_S$ of 0.230 at $\epsilon_B$ = 0.1. The performance is slightly lower for the ParT implementations.

The DeepSVDD method demonstrates the lowest AUC values among the techniques but still provides competitive results, especially for the MLP architecture with an AUC of 0.606 and a relatively high $\epsilon_S$ of 0.237 at $\epsilon_B$ = 0.1. The ParT architecture with DeepSVDD shows lower performance, with an AUC of 0.518 and $\epsilon_S$ of 0.103 at $\epsilon_B$ = 0.1, indicating that the simpler MLP architecture might be more suited for certain anomaly detection tasks when using DeepSVDD.

In addition, the performance of the anomaly detection method, namely the DDD ParT+SM model, was evaluated under different jet energy scale (JES) variations applied to the 4-top signal \cite{ATLAS:2014hvo,ATLAS:2017bje}. The results, presented in Appendix \ref{appendix:jes}, indicate that scaling the jet energies up or down introduced only minor variations in the anomaly detection metrics. This suggests that the model does not mistake systematic JES shifts for true anomalies and remains robust under realistic energy scale uncertainties. 
This study confirms that the JES variations applied to the 4-top signal do not mimic the characteristics of the generic anomaly the model is designed to detect.

These results suggest that while the ParT architecture generally outperforms the MLP in capturing complex correlations, the choice of anomaly detection technique plays a crucial role. The novel DDD method shows promising results and can be effectively combined with advanced architectures like ParT for improved anomaly detection.

\subsection{Comparison to the DarkMachines challenge algorithms}

Based on the significance improvement (SI) as defined in Ref.~\cite{Aarrestad_2022} as $\text{SI} = \epsilon_S/\sqrt{\epsilon_B}$, the total improvement (TI) was introduced to quantify the maximum SI across the various physics signals for each of the anomaly detection techniques and to combine the signals in multiple channels. This metric was used to analyse the minimum, median and maximum values for each of the physics models.

\begin{figure}
    \centering
    \begin{subfigure}[b]{1.\textwidth}
        \centering
        \includegraphics[width=\textwidth]{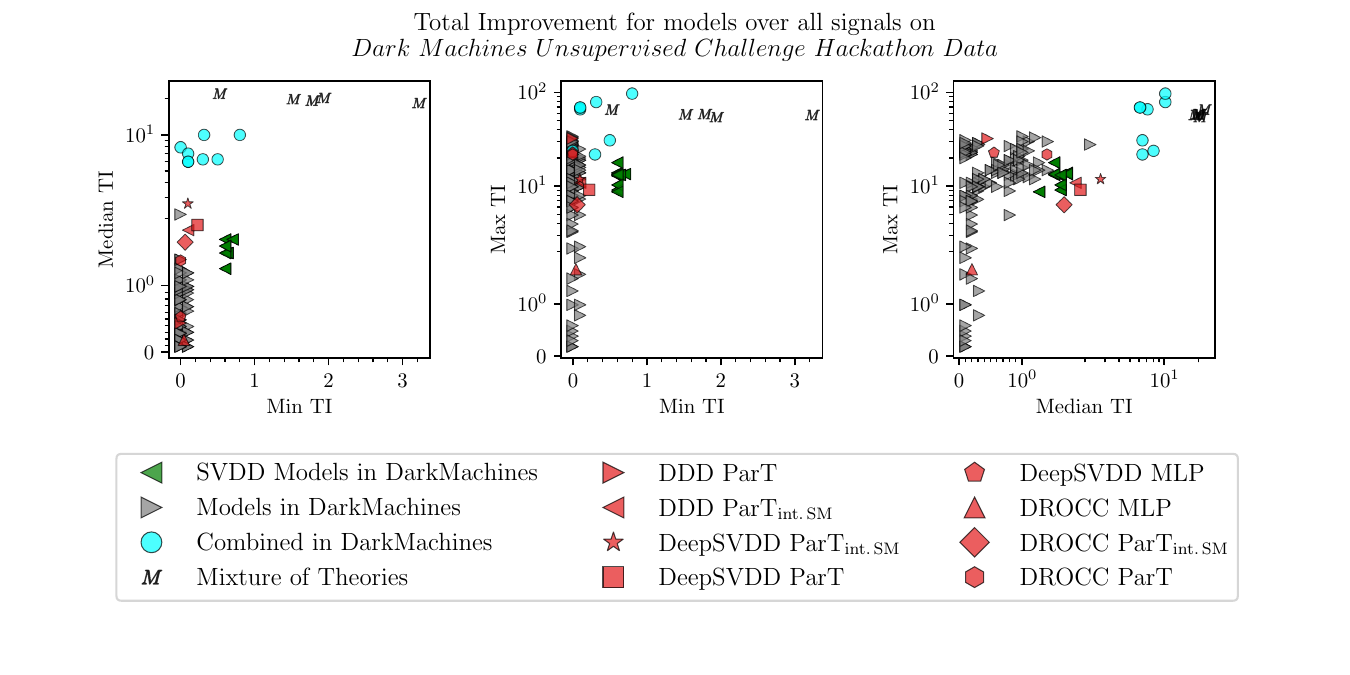}
    \end{subfigure}
    \caption{The minimum, median, and maximum best total improvements for each technique. 
    The metrics are computed with the description and channels of the DarkMachines challenge datasets.}
    \label{fig:darkmachines}
\end{figure}

Figure \ref{fig:darkmachines} provides a comparative analysis of the TI values for different models, including the results of different models in the DarkMachines challenge and the mixture of theories from Ref.~\cite{Caron_2023}. The TI is calculated over different background efficiency cuts ($\epsilon_B = 10^{-2}, 10^{-3},$ and $10^{-4}$), and the maximum is taken, which typically leads to $TI > 1$ values. 

The predictions show remarkably high median TI values, especially for the DeepSVDD models based on ParT architectures, both with and without the pairwise interaction matrix. The DDD method applied to the ParT with pairwise features also shows significant performance. In comparison, the models from the DarkMachines challenge also show high mean TI values. However, a direct comparison shows comparable results or slightly higher values in some cases, indicating that the methods presented in this work are competitive.

In terms of maximum TI, the models trained for this study, particularly the DeepSVDD and DDD with ParT architectures, reach a maximum TI of about 10. This is consistent with the best-performing models from the DarkMachines challenge, demonstrating the effectiveness of this approach. However, one notable difference is the minimum TI score. These models exhibit a lower minimum TI compared to the DarkMachines models. 

The performance across different channels in the DarkMachines challenge shows that models with a median TI greater than 2 are considered the best performing and advance to the next round. The models presented in this study, particularly those using the DeepSVDD and DDD techniques with the ParT architecture, show a median TI that qualifies (or almost classifies) them as top performers according to this criterion.
It should also be noted that the winning models in this challenge were combinations of SVDDs with normalising flow-based models (called `combined in DarkMachines' in Figure \ref{fig:darkmachines}). A combination with flow models would probably also further improve the performance on this ranking.
The $M$ models shown in the figure were created using a supervised classifier trained on a mixture of BSM signals, a method described in Ref.~\cite{Caron_2023}. This method could be superior to unsupervised methods if the signal has similar properties to the `BSM mixture'.

To summarize, the comparative analysis based on the median TI scores shows that these models, especially those using advanced architectures such as ParT with techniques such as DeepSVDD and DDD, perform quite well. They achieve competitive or even better results compared to the non-combined algorithms tested in the DarkMachines challenge, and overall the robustness and effectiveness of the models in detecting anomalies is demonstrated at the high mean and maximum TI values. This emphasizes the potential of turning state-of-the-art classifiers into anomaly detectors with minimal changes and using their capabilities to improve anomaly detection in high-energy physics experiments.

\pagebreak 

\section{Conclusions}
\label{sec: Conclusions}





This work demonstrates that a supervised classifier can be converted into an effective unsupervised anomaly detector. The proposed method, DDD, trained by discriminating between data with and without distortion, has shown promising results as an anomaly detector capable of identifying signals of new physics.
A comparison was made between the DDD method, DROCC (a semi-supervised method incorporating signals during training), and the DeepSVDD method. Results indicate that the effectiveness of each model depends on the specific signal and channel, with the proposed methods providing highly effective anomaly detection. 

Furthermore, it is noteworthy that the DDD model not only showed high effectiveness in detecting anomalies in different signals and channels among the evaluated methods, but also ranked among the three best performing models in the DarkMachines challenge.

For purely unsupervised applications, it is recommended to use a combination of DeepSVDD and DDD. For those with additional resources, integrating flow models can further enhance performance. These findings suggest that the best supervised models often excel in an unsupervised context as well, such as the particle transformer with SM interactions.

The study shows that it is likely possible to recognize 4-top production as an anomaly without prior knowledge of the process (/ as an anomaly without knowing the process beforehand).

Since the proposed anomaly detectors require only minimal changes to existing analysis systems, it is encouraged that the LHC community consider converting their supervised classifiers into anomaly detectors, in addition to supervised signal selection and signal fitting. It is suggested that every LHC search should include a search for anomalies.

\pagebreak

\section*{Acknowledgements}
The author(s) gratefully acknowledges the computer resources at Artemisa, funded by the European Union ERDF and Comunitat Valenciana as well as the technical support provided by the Instituto de Física Corpuscular, IFIC (CSIC-UV). The work of R. RdA was supported by PID2020-113644GB-I00 from the Spanish Ministerio de Ciencia e Innovación and by the PROMETEO/2022/69 from the Spanish GVA. The work of A.~R.~J. was supported by the project PID2021-124912NB-I00, funded by the Spanish Ministry MICIU. The work of M.~M.~Ll was supported by the RYC2019-028510-I grant (Spanish Ministry MICINN) and the ASFAE/2022/010 project (funded by Generalitat Valenciana and the European Union).

\pagebreak 
\appendix

\section*{Appendix}

This appendix supplements the main text with additional tables and analysis results. These tables provide a detailed summary of the performance metrics for various machine learning models and techniques applied to the DarkMachines challenge channels. Each table includes key metrics such as the area under the ROC curve (AUC), signal efficiency ($\epsilon_S$), and background efficiency ($\epsilon_B$).

\section{Performance Comparison Tables}



\subsection{Dark Machines channels} 
\label{appendix:dmachinesTables}

Tables~\ref{tab:ch1_all}, \ref{tab:ch2a_all}, \ref{tab:ch2b_all}, and \ref{tab:ch3_all} present detailed results for the Dark Machines channels. These include metrics like the AUC and background efficiencies for various signal processes. The information in these tables is used to evaluate and compare the performance of the methods and architectures discussed in the main text.

\begin{table}[H]
\centering
\resizebox{\textwidth}{!}{%
\setlength{\tabcolsep}{4pt}
\begin{tabular}{ll|ccc|cc|ccc}
\hline
\multicolumn{2}{c|}{} & \multicolumn{3}{c|}{DeepSVDD} & \multicolumn{2}{c|}{DDD} & \multicolumn{3}{c}{DROCC} \\
 & & MLP & ParT & $\text{ParT}_{\text{SM}}$ & ParT & $\text{ParT}_{\text{SM}}$ & MLP & ParT & $\text{ParT}_{\text{SM}}$ \\ 
\hline \hline
\multirow{3}{*}{glgl1400} & AUC & 0.560 & 0.690 & 0.670 & 0.726 & \bf{0.822} & 0.652 & 0.755 & 0.762 \\
& $\epsilon_B(\epsilon_S=0.01)$ & 0.023 & 0.001 & 0.006 & 0.003 & \boldmath{$<0.001$} & 0.006 & 0.001 & 0.001 \\
& $\epsilon_S(\epsilon_B=0.01)$ & 0.003 & 0.076 & 0.023 & 0.032 & \bf{0.172} & 0.032 & 0.084 & 0.070 \\
\hline
\multirow{3}{*}{glgl1600} & AUC & 0.590 & 0.850 & 0.820 & 0.801 & 0.895 & 0.728 & 0.828 & \bf{0.904} \\
& $\epsilon_B(\epsilon_S=0.01)$ & 0.023 & \boldmath{$<0.001$} & 0.003 & \boldmath{$<0.001$} & \boldmath{$<0.001$} & 0.004 & \boldmath{$<0.001$} & \boldmath{$<0.001$} \\
& $\epsilon_S(\epsilon_B=0.01)$ & 0.004 & 0.224 & 0.066 & 0.129 & \bf{0.333} & 0.064 & 0.16 & 0.306 \\
\hline 
\multirow{3}{*}{monojet} & AUC & 0.520 & 0.840 & 0.810 & 0.513 & 0.769 & 0.611 & 0.734 & \bf{0.844} \\
& $\epsilon_B(\epsilon_S=0.01)$ & 0.048 & \boldmath{$<0.001$} & 0.002 & 0.012 & 0.002 & 0.008 & 0.002 & \boldmath{$<0.001$} \\
& $\epsilon_S(\epsilon_B=0.01)$ & 0.003 & \bf{0.270} & 0.108 & 0.009 & 0.079 & 0.033 & 0.070 & 0.100 \\
\hline
\multirow{3}{*}{monotop} & AUC & 0.720 & \bf{0.740} & \bf{0.740} & 0.580 & 0.709 & 0.596 & 0.679 & 0.733 \\
& $\epsilon_B(\epsilon_S=0.01)$ & 0.003 & \boldmath{$<0.001$} & 0.001 & 0.007 & 0.008 & 0.015 & 0.002 & 0.001 \\
& $\epsilon_S(\epsilon_B=0.01)$ & 0.030 & \bf{0.125} & 0.062 & 0.016 & 0.034 & 0.016 & 0.042 & 0.062 \\
\hline
\multirow{3}{*}{sqsq1} & AUC & 0.680 & 0.770 & 0.780 & 0.786 & \bf{0.850} & 0.746 & 0.800 & 0.830 \\
& $\epsilon_B(\epsilon_S=0.01)$ & 0.001 & 0.001 & 0.001 & \boldmath{$<0.001$} & 0.004 & 0.006 & \boldmath{$<0.001$} & \boldmath{$<0.001$} \\
& $\epsilon_S(\epsilon_B=0.01)$ & 0.086 & 0.099 & 0.082 & 0.140 & 0.055 & 0.051 & 0.108 & \bf{0.161} \\
\hline
\multirow{3}{*}{sqsq} & AUC & 0.590 & 0.950 & 0.930 & 0.699 & 0.884 & 0.794 & 0.852 & \bf{0.955} \\
& $\epsilon_B(\epsilon_S=0.01)$ & 0.031 & \boldmath{$<0.001$} & 0.001 & 0.001 & \boldmath{$<0.001$} & 0.006 & 0.001 & \boldmath{$<0.001$} \\
& $\epsilon_S(\epsilon_B=0.01)$ & 0.001 & \bf{0.600} & 0.350 & 0.078 & 0.314 & 0.102 & 0.219 & 0.486 \\
\hline
\multirow{3}{*}{stlp} & AUC & \bf{0.960} & 0.930 & \bf{0.960} & 0.749 & 0.767 & 0.920 & 0.848 & 0.896 \\
& $\epsilon_B(\epsilon_S=0.01)$ & \boldmath{$<0.001$} & \boldmath{$<0.001$} & \boldmath{$<0.001$} & \boldmath{$<0.001$} & 0.026 & 0.001 & 0.293 & \boldmath{$<0.001$} \\
& $\epsilon_S(\epsilon_B=0.01)$ & 0.419 & 0.360 & \bf{0.537} & 0.286 & 0.002 & $<0.001$ & 0.078 & 0.140 \\
\hline
\multirow{3}{*}{stop2b} & AUC & 0.830 & 0.870 & 0.860 & 0.764 & 0.861 & 0.735 & 0.843 & \bf{0.887} \\
& $\epsilon_B(\epsilon_S=0.01)$ & 0.001 & \boldmath{$<0.001$} & 0.001 & 0.001 & 0.008 & 0.018 & 0.001 & \boldmath{$<0.001$} \\
& $\epsilon_S(\epsilon_B=0.01)$ & 0.084 & 0.231 & 0.102 & 0.112 & 0.023 & 0.040 & 0.168 & \bf{0.239} \\
\hline  
\end{tabular}%
}
\caption{Channel 1 results. The metrics for the SVDD and DDD methods are calculated based on the average predictions of four ensembled models. Detailed information about the calculation is provided in the main text. Bold numbers represent the highest performance. When multiple architectures achieve the best performance, all corresponding results are highlighted.}
\label{tab:ch1_all}
\end{table}

\begin{table}[H]
\centering
\resizebox{\textwidth}{!}{%
\setlength{\tabcolsep}{4pt}
\begin{tabular}{ll|ccc|cc|ccc}
\hline
\multicolumn{2}{c|}{} & \multicolumn{3}{c|}{DeepSVDD} & \multicolumn{2}{c|}{DDD} & \multicolumn{3}{c}{DROCC}  \\
 & & MLP & ParT & $\text{ParT}_{\text{SM}}$ & ParT & $\text{ParT}_{\text{SM}}$ & MLP & ParT & $\text{ParT}_{\text{SM}}$ \\
\hline \hline
\multirow{3}{*}{cha200} & AUC    & 0.407 & \bf{0.569} & 0.555 & 0.315 & 0.464 & 0.397 & 0.412 & 0.487 \\
& $\epsilon_B(\epsilon_S=0.01)$  & 0.382 & \bf{0.004} & 0.012 & 0.014 & 0.014 & 0.009 & 0.008 & 0.006 \\
& $\epsilon_S(\epsilon_B=0.01)$  & $<0.001$ & 0.014 & 0.005 & 0.010 & 0.005 & 0.001 & 0.014 & \bf{0.023} \\
\hline
\multirow{3}{*}{cha250} & AUC    & 0.382 & 0.531 & \bf{0.571} & 0.330 & 0.427 & 0.410 & 0.400 & 0.450 \\
& $\epsilon_B(\epsilon_S=0.01)$  & 0.033 & \bf{0.008} & 0.009 & 0.031 & 0.013 & 0.009 & 0.042 & 0.035 \\
& $\epsilon_S(\epsilon_B=0.01)$  & $<0.001$ & \bf{0.022} & 0.013 & 0.005 & 0.006 & 0.005 & 0.001 & 0.006 \\
\hline
\multirow{3}{*}{cha300} & AUC    & 0.440 & 0.600 & \bf{0.601} & 0.328 & 0.450 & 0.369 & 0.394 & 0.528 \\
& $\epsilon_B(\epsilon_S=0.01)$  & 0.060 & \bf{0.007} & 0.010 & 0.039 & 0.013 & 0.022 & 0.010 & 0.022 \\
& $\epsilon_S(\epsilon_B=0.01)$  & $<0.001$ & \bf{0.023} & 0.010 & 0.003 & 0.010 & 0.003 & 0.009 & 0.006 \\
\hline
\multirow{3}{*}{cha400} & AUC    & 0.445 & 0.578 & \bf{0.582} & 0.334 & 0.433 & 0.365 & 0.402 & 0.510 \\
& $\epsilon_B(\epsilon_S=0.01)$  & 0.058 & 0.005 & \bf{0.003} & 0.031 & 0.025 & 0.031 & 0.017 & 0.041 \\
& $\epsilon_S(\epsilon_B=0.01)$  & $<0.001$ & 0.022 & \bf{0.031} & 0.003 & 0.002 & 0.003 & 0.006 & 0.006 \\
\hline
\multirow{3}{*}{pp23mt} & AUC    & 0.378 & 0.406 & 0.485 & 0.385 & 0.557 & 0.604 & \bf{0.628} & 0.593 \\
& $\epsilon_B(\epsilon_S=0.01)$  & 0.034 & 0.013 & 0.015 & 0.006 & 0.005 & \boldmath{$<0.001$} & \boldmath{$<0.001$} & 0.023 \\
& $\epsilon_S(\epsilon_B=0.01)$  & 0.004 & 0.009 & 0.003 & 0.015 & 0.027 & 0.036 & \bf{0.175} & 0.004 \\
\hline
\multirow{3}{*}{pp24mt} & AUC    & 0.529 & 0.531 & 0.539 & 0.633 & 0.687 & \bf{0.697} & 0.663 & 0.604 \\
& $\epsilon_B(\epsilon_S=0.01)$  & 0.033 & 0.005 & 0.008 & 0.001 & 0.010 & 0.001 & \boldmath{$<0.001$} & 0.022 \\
& $\epsilon_S(\epsilon_B=0.01)$  & 0.003 & 0.016 & 0.016 & 0.060 & 0.011 & 0.061 & \bf{0.193} & 0.004 \\
\hline
\multirow{3}{*}{gluino} & AUC    & \bf{0.960} & 0.794 & 0.719 & 0.915 & 0.611 & 0.893 & 0.745 & 0.783 \\
& $\epsilon_B(\epsilon_S=0.01)$  & \boldmath{$<0.001$} & \boldmath{$<0.001$} & 0.001 & \boldmath{$<0.001$} & \boldmath{$<0.001$} & 0.001 & 0.003 & 0.001 \\
& $\epsilon_S(\epsilon_B=0.01)$  & 0.562 & 0.087 & 0.050 & \bf{0.661} & 0.046 & 0.099 & 0.067 & 0.153 \\
\hline
\end{tabular}%
}
\caption{Channel 2a results. The metrics for the SVDD and DDD methods are calculated based on the average predictions of four ensembled models. Detailed information about the calculation is provided in the main text. Bold numbers represent the highest performance. When multiple architectures achieve the best performance, all corresponding results are highlighted.}
\label{tab:ch2a_all}
\end{table}


\begin{table}[H]
\centering
\resizebox{\textwidth}{!}{%
\setlength{\tabcolsep}{4pt}
\begin{tabular}{ll|ccc|cc|ccc}
\hline
\multicolumn{2}{c|}{} & \multicolumn{3}{c|}{DeepSVDD} & \multicolumn{2}{c|}{DDD} & \multicolumn{3}{c}{DROCC}  \\
 & & MLP & ParT & $\text{ParT}_{\text{SM}}$ & ParT & $\text{ParT}_{\text{SM}}$ & MLP & ParT & $\text{ParT}_{\text{SM}}$ \\
\hline \hline
\multirow{3}{*}{cha300} & AUC & 0.620 & \bf{0.770} & 0.760 & 0.493 & 0.717 & 0.585 & 0.578 & 0.595 \\
& $\epsilon_B(\epsilon_S=0.01)$ & 0.008 & 0.002 & 0.001 & 0.006 & \boldmath{$<0.001$} & 0.001 & 0.019 & 0.028 \\
& $\epsilon_S(\epsilon_B=0.01)$ & 0.014 & 0.080 & \bf{0.105} & 0.013 & 0.095 & 0.051 & 0.018 & 0.018 \\
\hline
\multirow{3}{*}{cha400} & AUC & 0.700 & 0.790 & \bf{0.800} & 0.522 & 0.730 & 0.654 & 0.645 & 0.637 \\
& $\epsilon_B(\epsilon_S=0.01)$ & 0.005 & 0.001 & \boldmath{$<0.001$} & 0.002 & \boldmath{$<0.001$} & 0.002 & 0.013 & 0.027 \\
& $\epsilon_S(\epsilon_B=0.01)$ & 0.015 & 0.115 & \bf{0.167} & 0.036 & 0.068 & 0.049 & 0.019 & 0.051 \\
\hline 
\multirow{3}{*}{cha600} & AUC & 0.730 & 0.820 & \bf{0.830} & 0.503 & 0.720 & 0.684 & 0.704 & 0.672 \\
& $\epsilon_B(\epsilon_S=0.01)$ & 0.004 & 0.001 & \boldmath{$<0.001$} & 0.006 & 0.001 & 0.002 & 0.009 & 0.019 \\
& $\epsilon_S(\epsilon_B=0.01)$ & 0.014 & 0.164 & \bf{0.211} & 0.023 & 0.077 & 0.056 & 0.032 & 0.093 \\
\hline
\multirow{3}{*}{cha200} & AUC & 0.610 & \bf{0.690} & 0.670 & 0.576 & 0.644 & 0.626 & 0.563 & 0.592 \\
& $\epsilon_B(\epsilon_S=0.01)$ & \boldmath{$<0.001$} & 0.001 & \boldmath{$<0.001$} & 0.003 & \boldmath{$<0.001$} & 0.002 & 0.012 & 0.016 \\
& $\epsilon_S(\epsilon_B=0.01)$ & 0.059 & 0.098 & \bf{0.115} & 0.023 & 0.051 & 0.056 & 0.011 & 0.044 \\
\hline  
\multirow{3}{*}{cha250} & AUC & 0.580 & 0.640 & 0.620 & 0.594 & \bf{0.649} & 0.619 & 0.524 & 0.572 \\
& $\epsilon_B(\epsilon_S=0.01)$ & 0.001 & \boldmath{$<0.001$} & \boldmath{$<0.001$} & 0.004 & 0.001 & 0.002 & 0.017 & 0.016 \\
& $\epsilon_S(\epsilon_B=0.01)$ & 0.070 & 0.108 & \bf{0.120} & 0.018 & 0.056 & 0.042 & 0.011 & 0.051 \\
\hline  
\multirow{3}{*}{gluino} & AUC & 0.985 & 0.890 & 0.900 & \bf{0.989} & 0.913 & 0.943 & 0.934 & 0.827 \\
& $\epsilon_B(\epsilon_S=0.01)$ & \boldmath{$<0.001$} & 0.001 & 0.001 & \boldmath{$<0.001$} & \boldmath{$<0.001$} & 0.003 & \boldmath{$<0.001$} & 0.006 \\
& $\epsilon_S(\epsilon_B=0.01)$ & 0.701 & 0.134 & 0.182 & \bf{0.880} & 0.464 & 0.102 & 0.303 & 0.234 \\
\hline  
\multirow{3}{*}{pp23mt} & AUC & 0.650 & 0.740 & 0.730 & 0.699 & \bf{0.903} & 0.658 & 0.44 & 0.551 \\
& $\epsilon_B(\epsilon_S=0.01)$ & \boldmath{$<0.001$} & 0.001 & \boldmath{$<0.001$} & 0.002 & \boldmath{$<0.001$} & 0.001 & 0.022 & 0.015 \\
& $\epsilon_S(\epsilon_B=0.01)$ & 0.283 & 0.248 & 0.282 & 0.029 & \bf{0.374} & 0.059 & 0.009 & 0.074 \\
\hline  
\multirow{3}{*}{pp24mt} & AUC & 0.720 & 0.760 & 0.770 & 0.737 & \bf{0.923} & 0.725 & 0.484 & 0.602 \\
& $\epsilon_B(\epsilon_S=0.01)$ & \boldmath{$<0.001$} & \boldmath{$<0.001$} & \boldmath{$<0.001$} & 0.003 & \boldmath{$<0.001$} & \boldmath{$<0.001$} & 0.014 & 0.009 \\
& $\epsilon_S(\epsilon_B=0.01)$ & 0.368 & 0.359 & \bf{0.372} & 0.024 & 0.293 & 0.074 & 0.009 & 0.074 \\
\hline  
\multirow{3}{*}{stlp} & AUC & \bf{0.960} & 0.880 & 0.900 & 0.827 & 0.751 & 0.878 & 0.904 & 0.834 \\
& $\epsilon_B(\epsilon_S=0.01)$ & \boldmath{$<0.001$} & 0.001 & \boldmath{$<0.001$} & 0.005 & \boldmath{$<0.001$} & 0.002 & 0.001 & 0.003 \\
& $\epsilon_S(\epsilon_B=0.01)$ & 0.184 & 0.200 & \bf{0.332} & 0.030 & 0.101 & 0.208 & 0.126 & 0.126 \\
\hline   
\end{tabular}%
}
\caption{Channel 2b results. The metrics for the SVDD and DDD methods are calculated based on the average predictions of four ensembled models. Detailed information about the calculation is provided in the main text. Bold numbers represent the highest performance. When multiple architectures achieve the best performance, all corresponding results are highlighted.}
\label{tab:ch2b_all}
\end{table}

\begin{table}[H]
\centering
\resizebox{\textwidth}{!}{%
\setlength{\tabcolsep}{4pt}
\begin{tabular}{ll|ccc|cc|ccc}
\hline
\multicolumn{2}{c|}{} & \multicolumn{3}{c|}{DeepSVDD} & \multicolumn{2}{c|}{DDD} & \multicolumn{3}{c}{DROCC}  \\
 & & MLP & ParT & $\text{ParT}_{\text{SM}}$ & ParT & $\text{ParT}_{\text{SM}}$ & MLP & ParT & $\text{ParT}_{\text{SM}}$ \\
\hline \hline
\multirow{3}{*}{glgl1400} & AUC   & 0.712 & 0.755 & 0.786 & 0.819 & \bf{0.824} & 0.592 & 0.811 & 0.714 \\
& $\epsilon_B(\epsilon_S=0.01)$  & 0.013 & 0.003 & 0.001 & \boldmath{$<0.001$} & \boldmath{$<0.001$} & 0.017 & 0.002 & 0.001 \\
& $\epsilon_S(\epsilon_B=0.01)$  & 0.004 & 0.063 & 0.159 & 0.167 & 0.163 & 0.003 & \bf{0.187} & 0.098 \\
\hline
\multirow{3}{*}{glgl1600} & AUC    & 0.779 & 0.883 & \bf{0.909} & 0.808 & 0.806 & 0.638 & 0.841 & 0.804 \\
& $\epsilon_B(\epsilon_S=0.01)$  & 0.010 & 0.002 & \boldmath{$<0.001$} & \boldmath{$<0.001$} & \boldmath{$<0.001$} & 0.015 & 0.001 & 0.003 \\
& $\epsilon_S(\epsilon_B=0.01)$  & 0.009 & 0.276 & \bf{0.456} & 0.168 & 0.172 & 0.003 & 0.223 & 0.181 \\
\hline
\multirow{3}{*}{gluino} & AUC    & 0.965 & 0.898 & 0.897 & 0.966 & \bf{0.974} & 0.929 & 0.930 & 0.731 \\
& $\epsilon_B(\epsilon_S=0.01)$  & \boldmath{$<0.001$} & 0.001 & 0.001 & \boldmath{$<0.001$} & \boldmath{$<0.001$} & 0.002 & 0.001 & 0.122 \\
& $\epsilon_S(\epsilon_B=0.01)$  & 0.565 & 0.170 & 0.142 & 0.616 & \bf{0.635} & 0.168 & 0.634 & 0.277 \\
\hline
\multirow{3}{*}{monojet} & AUC    & 0.650 & 0.945 & \bf{0.954} & 0.384 & 0.445 & 0.502 & 0.565 & 0.827 \\
& $\epsilon_B(\epsilon_S=0.01)$  & 0.011 & \boldmath{$<0.001$} & \boldmath{$<0.001$} & 0.062 & 0.003 & 0.021 & 0.042 & 0.001 \\
& $\epsilon_S(\epsilon_B=0.01)$  & 0.005 & 0.469 & \bf{0.645} & $<0.001$ & 0.014 & 0.004 & 0.001 & 0.277 \\
\hline
\multirow{3}{*}{monotop} & AUC    & 0.719 & 0.819 & \bf{0.827} & 0.589 & 0.599 & 0.564 & 0.686 & 0.712 \\
& $\epsilon_B(\epsilon_S=0.01)$  & 0.001 & \boldmath{$<0.001$} & \boldmath{$<0.001$} & 0.011 & 0.006 & 0.021 & 0.004 & 0.001 \\
& $\epsilon_S(\epsilon_B=0.01)$  & 0.083 & 0.191 & \bf{0.270} & 0.009 & 0.015 & 0.003 & 0.042 & 0.086 \\
\hline
\multirow{3}{*}{monoV} & AUC    & 0.677 & 0.943 & \bf{0.952} & 0.351 & 0.448 & 0.522 & 0.525 & 0.816 \\
& $\epsilon_B(\epsilon_S=0.01)$  & 0.001 & \boldmath{$<0.001$} & \boldmath{$<0.001$} & 0.099 & 0.004 & 0.016 & 0.081 & \boldmath{$<0.001$} \\
& $\epsilon_S(\epsilon_B=0.01)$  & 0.055 & 0.486 & \bf{0.642} & $<0.001$ & 0.014 & 0.004 & $<0.001$ & 0.223 \\
\hline
\multirow{3}{*}{sqsq1} & AUC    & 0.821 & 0.856 & \bf{0.877} & 0.844 & 0.842 & 0.702 & 0.845 & 0.751 \\
& $\epsilon_B(\epsilon_S=0.01)$  & \boldmath{$<0.001$} & 0.001 & \boldmath{$<0.001$} & \boldmath{$<0.001$} & \boldmath{$<0.001$} & 0.011 & \boldmath{$<0.001$} & \boldmath{$<0.001$} \\
& $\epsilon_S(\epsilon_B=0.01)$  & 0.232 & 0.212 & \bf{0.282} & 0.203 & 0.184 & 0.009 & 0.278 & 0.120 \\
\hline
\multirow{3}{*}{sqsq} & AUC    & 0.726 & 0.959 & \bf{0.965} & 0.540 & 0.546 & 0.621 & 0.713 & 0.867 \\
& $\epsilon_B(\epsilon_S=0.01)$  & 0.011 & 0.001 & \boldmath{$<0.001$} & 0.011 & 0.002 & 0.018 & 0.003 & \boldmath{$<0.001$} \\
& $\epsilon_S(\epsilon_B=0.01)$  & 0.007 & 0.661 & \bf{0.759} & 0.009 & 0.019 & 0.003 & 0.041 & 0.288 \\
\hline
\multirow{3}{*}{stlp} & AUC    & 0.962 & \bf{0.963} & \bf{0.963} & 0.774 & 0.786 & 0.621 & 0.930 & 0.731 \\
& $\epsilon_B(\epsilon_S=0.01)$  & \boldmath{$<0.001$} & \boldmath{$<0.001$} & \boldmath{$<0.001$} & \boldmath{$<0.001$} & \boldmath{$<0.001$} & 0.002 & \boldmath{$<0.001$} & 0.122 \\
& $\epsilon_S(\epsilon_B=0.01)$  & \bf{0.830} & 0.682 & 0.635 & 0.137 & 0.143 & 0.003 & 0.634 & 0.277 \\
\hline
\multirow{3}{*}{st2b} & AUC    & 0.860 & 0.913 & \bf{0.920} & 0.830 & 0.841 & 0.726 & 0.876 & 0.798 \\
& $\epsilon_B(\epsilon_S=0.01)$  & \boldmath{$<0.001$} & 0.001 & \boldmath{$<0.001$} & \boldmath{$<0.001$} & \boldmath{$<0.001$} & 0.012 & \boldmath{$<0.001$} & \boldmath{$<0.001$} \\
& $\epsilon_S(\epsilon_B=0.01)$  & 0.195 & 0.315 & \bf{0.417} & 0.168 & 0.169 & 0.006 & 0.337 & 0.186 \\
\hline
\end{tabular}%
}
\caption{Channel 3 results. The metrics for the SVDD and DDD methods are calculated based on the average predictions of four ensembled models. Detailed information about the calculation is provided in the main text. Bold numbers represent the highest performance. When multiple architectures achieve the best performance, all corresponding results are highlighted.}
\label{tab:ch3_all}
\end{table}

\section{JES studies}
\label{appendix:jes}

To investigate the potential impact of jet energy scale (JES) variations on the performance of the DDD ParT+SM model, a study was conducted in which the jet energy of signal (4-top) events was systematically scaled up and down by different percentages. The aim was to examine whether these systematic variations in jet energy could significantly affect anomaly detection results.

Fixed scaling factors of $\pm 2\%$, $\pm 3\%$, $\pm 5\%$, $\pm 8\%$, and $\pm 10\%$ were applied to the jet and b-jet energies, with the transverse momentum ($p_T$) recalculated accordingly. 

\begin{figure}[h!]
    \centering
    \begin{minipage}{0.90\textwidth}
        \centering
        \includegraphics[width=\textwidth]{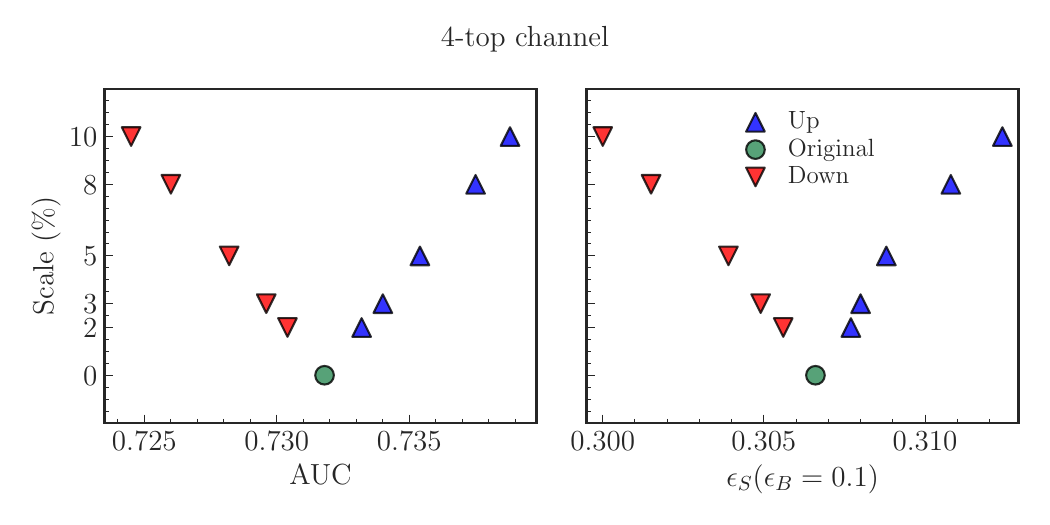}
    \end{minipage}
    \caption{Performance of the DDD ParT+SM in the 4-top channel with JES variations of up to 10\%. The left plot shows AUC values, while the right plot displays signal efficiency ($\epsilon_S$) for a fixed background efficiency of 10\%. The markers represent different scale types: `Up' (blue) and `Down' (red).  The green marker corresponds to the original model without any applied scaling.}
    \label{fig: scaling}
\end{figure}

The results, presented in Fig.~\ref{fig: scaling}, demonstrate that while scaling jet energies up or down introduces minor systematic changes in AUC and signal efficiency ($\varepsilon_S$), these variations remain relatively small and gradual across all tested scale percentages. This suggests that the model remains stable under reasonable JES variations.

In Section~\ref{subsec: 4top}, the results of this study are discussed in the context of assessing the robustness of the DDD method when systematic uncertainties, such as JES, are considered.

\clearpage

\clearpage
\bibliographystyle{JHEP}
\bibliography{refs.bib}

\end{document}